# Acetylene – Argon Plasmas Measured at a Biased Substrate Electrode for Diamond-Like Carbon Deposition, Part 2: Ion Energy Distributions.

**A Baby, C M O Mahony, P. Lemoine and P D Maguire**

Nanotechnology and Integrated Bio-Engineering Centre (NIBEC), University of Ulster, Newtownabbey, BT37 0QB, N. Ireland

**Abstract.** Ion energy distributions have been determined at the rf-bias electrode in an inductively-coupled acetylene-argon plasma for various substrate bias voltages and frequencies under conditions suitable for film deposition. These are compared with those obtained at the grounded wall of a capacitively coupled plasma. In the former, for pressures up to 25 mTorr, the IEDs exhibit bimodal structures with peak separation values that follow the expected dependence on voltage and frequency. At higher pressures, 120 mTorr the bimodal structure is replaced by a single peak. For all conditions the dominant ion is $Ar^+$ or $ArH^+$ despite the set flow ratio of $C_2H_2$:Ar of 2:1 and this can be attributed to the high electron dissociation of the parent molecule. Diamond-like carbon films indicate a peak hardness at an ion energy of around 90 eV and a very sharp fall in hardness is noted beyond this value. This is similar to observed $sp^3$ bond formation in hydrogen-free tetragonal amorphous carbon or bias-sputtered films. However, due to the lack of carbon-based ions, an alternative mechanism is likely, based on argon knock-on implantation of surface adsorbed carbon species. The results have shown that use of high frequency bias or bias harmonics may lead to much narrower ion energy distributions.

**1. Introduction**

Diamond-like or amorphous carbons (DLC) represent a class of technologically important thin film materials. The ability to vary properties such as hardness, Young's modulus, surface roughness, electrical resistance, thermal conductivity, density, refractive index, among others, offers considerable versatility in mechanical, electrical, optical and more recently biomedical applications. Intensive study over the past decade, using a wide range of complementary high resolution analysis techniques and atomistic modelling has resulted in detailed understanding of the relationship between functional properties and material structure. Nevertheless this work has been most successful in the case of hydrogen-free tetragonally-bonded amorphous carbon (t-aC). Filtered ion beams (mass-selected or filtered cathodic arc) are generally used for t-aC deposition and thus the relationship between deposition conditions and material properties can be readily established. In particular it is accepted that within an ion energy window of approximately 50 eV – 100 eV, energetic species subplantation and $sp^3$ carbon-bond formation is favoured over high energy relaxation to graphitic $sp^2$-bonding resulting in diamond-like ultra-hard and smooth thin films. Unfortunately issues such as cost, plant complexity, high intrinsic stress and line of sight planar deposition restrict the use of t-aC to high end specialist applications.

Hydrogenated amorphous carbon (a-C:H), although it does not approach the quality of t-aC, is suitable for many applications and can be routinely deposited in simple capacitively-coupled radio frequency (rf) plasma enhanced chemical vapour deposition (PECVD) systems. Hence it is widely favoured. Research has investigated factors such as power, bias, pressure, hydrocarbon precursors and inert gas dilution in order to optimise processes for a given application. More detailed plasma studies are seriously hampered by instrument and probe contamination and the species-rich plasma environment limits our ability to derive accurate relationships between structure, function and deposition conditions. From the limited number of mass spectrometry and ion energy studies, we understand in general that energetic hydrocarbon ions bombarding a surface play a major role in a-C:H formation. These act in concert with radical adsorption; the particular growth pathway being dependent upon the dominant radical which is in turn depends upon choice of pre-cursor gas (e.g. $CH_4$, $C_2H_2$ etc.), pressure and its dilution (e.g. Ar). Jacob [1] and co-workers [2] have highlighted, through extensive beam experiments coupled with complicated in-situ materials diagnostics, the highly complex species interactions at play, both synergistic and "anti-synergistic". The interplay between ion-bombardment (inert and reactive), radical adsorption and desorption as well as hydrogen reactions via molecular dissociation, adsorption/desorption, bond-formation and etching is modelled in detail but transferring these model outcomes to applications processing in standard systems is still problematic. Overall, from a practical applications perspective, the ability to predict and tailor a-C:H properties is limited. In fact, it is accepted that in standard PECVD, the a-C:H properties are relatively insensitive to plasma input variables. At very low powers or grounded electrodes, the low self-bias results in soft polymeric a-C:H with a large bandgap, while at very high powers (bias > 500 V), the graphitic phase dominates. A number of other, more specialist, plasma sources have been reported including ICP, ECR [1] and cascaded arc [3] and found to be viable. In particular, these systems offer higher deposition rates due to higher plasma densities achievable at low to moderate substrate biases. Future directions in a-C:H processing include accommodating complex and 3-dimensional substrates for biomedical and other applications, large area coatings (e.g. packaging, barrier and anti-corrosion layers) and a wider range of substrate materials (e.g. polymers, non-carbide forming metals).

The aim of this work therefore is to investigate the use of in-situ plasma diagnostics under realistic DLC process conditions, to gain an understanding of the plasma conditions that have significance in DLC deposition and to explore options to expand the process parameter space. To this end we have used mass spectrometry and mass-selected ion spectrometry at the rf-bias substrate electrode to determine neutral and ion fluxes in an ICP reactor, along with infra-red absorption studies for temporal neutral species evolution after argon-acetylene plasma initiation. These results are presented in an associated paper [4]. In this paper, ion energy distributions (IED) for $Ar^+$ and the dominant $H_xC_y^+$ species are determined at the biased substrate for various values of mean rf bias, bias frequency, pressure and $Ar/C_2H_2$ flow ratios. For comparison, IEDs are also measured in a capacitively-coupled plasma system that has been routinely used for diamond-like carbon deposition. In this case, access for the mass-energy spectrometer is only available at the grounded chamber wall. Species flux data is compared with previously reported spectra taken at remote locations from the substrate for use in subsequent plasma models. Deposited film properties are evaluated under a range of bias and pressure conditions in order to derive guiding principles for process optimisation in terms of mean ion energy and ion energy spread and their relationship to bias voltage and frequency, pressure and gas flow ratio. Film characteristics are compared with those obtained by other techniques (CCP, cathodic arc and sputtering) in order to gain insight into possible growth mechanism.

## 2. Experimental

The measurements described in this study were performed in two rf driven plasma facilities, one employing a Capacitively Coupled Plasma (CCP) and the other an Inductively Coupled Plasma (ICP). The working gases were usually pure argon or 2:1 $C_2H_2$:Ar mixtures; the small number of measurements made with other flow ratios are noted in the text.

### 2.1 CCP

The CCP comprised a 360 mm diameter 480 mm tall cylindrical chamber attached to associated gas, vacuum and electrical supplies. Gas flow controllers were used to set the flow rates (and thus flow ratios) of the working gases over the pressure range studied. A 13.56 MHz rf power supply and matching network was used to apply an rf potential to the 270 m diameter driven (substrate) electrode, sited centrally at the base of the chamber. To measure the ion energy distribution, the chamber was adapted to house a HIDEN EQP 300 mass energy analyser (MEA), which was sited at the side wall of the chamber, 170 mm above the top face of the driven electrode. The analyser was operated in Ion Energy (IE) mode to give the Ion Energy Distribution (IED). The measurements were made at 12 mTorr CCP dc bias ($V_{dc}$) of 450 V.

### 2.2 ICP

The ICP chamber (figure 1) is a 400 mm diameter sphere with six orthogonal 250 mm diameter ports which provide access for plasma systems and diagnostics. Needle valves were used to set the flow rates and ratios of the working gases within a pressure range of 3.3 mTorr to 120 mTorr. The 13.56 MHz rf power supply and close coupled automatic matching network drives an rf current through the water cooled 152 mm diameter flat spiral ICP coil, which has four lobes to inhibit E-mode operation. The oscillating rf current in the coil couples to the plasma via a 12.5 mm thick quartz window below which a 120 mm diameter substrate electrode is sited. The window to electrode separation was 160 mm for all measurements. A second, variable frequency ($f_{bias}$), rf power supply and matching network was used to apply a negative bias ($V_{bias}$) to the substrate electrode. The substrate electrode was coupled, via a central 50 μm diameter hole, to the MEA, which was operated in Ion Energy (IE) mode. A Scientific Systems compensated Langmuir probe was used to measure plasma parameters 107 mm above the bias electrode, at a radius of 45 mm from the vertical axis.

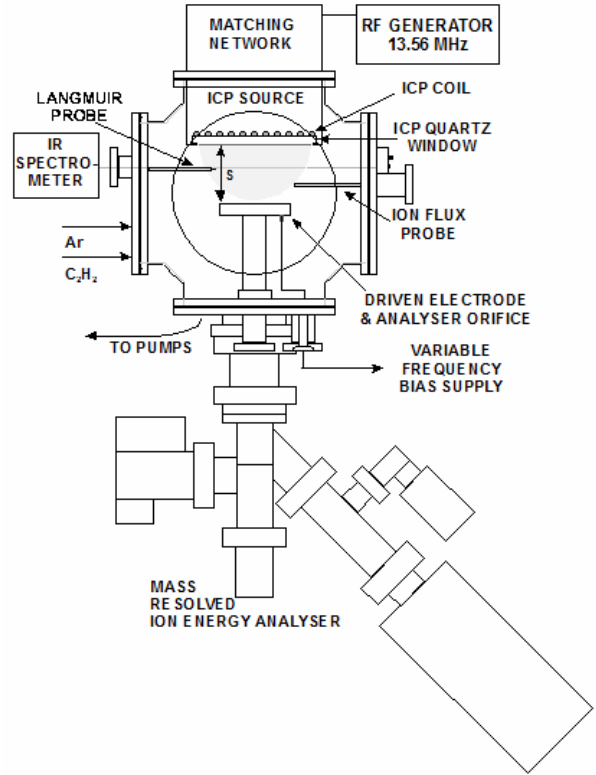

Figure 1 : ICP chamber and diagnostics schematic diagram.

This study concentrates on the ion energy distribution at the biased electrode with varying pressure, $V_{bias}$, $f_{bias}$ and on DLC deposition at 3.3 mTorr. In most cases the ICP input power was set at 200 W and automatically matched so that the reflected power was < 1 W. The working gases were again pure argon or 2:1 $C_2H_2$:Ar.

DLC (a-C:H) films were deposited on silicon samples placed on the rf biased electrode. Samples (10 x 10 $mm^2$) were cleaned using deionised water and ultrasonic treatment before loading into the vacuum chamber. The samples were further cleaned using $Ar^+$ ion sputter at 3.3 mTorr and 100 V bias voltage, prior to deposition. The depositions were performed at 3.3 mTorr working pressure, for a range of bias voltages 7 V to 100 V at 8.311 MHz bias frequency using 33% argon in $C_2H_2$:Ar. Film thicknesses were measured using a Dektak stylus profilometer and surface characterisation was done using ISA LabRam Raman and Nanoindentation.

## 3. Results
### 3.1 Ion Energy Distribution

For the CCP configuration, IEDs of the main positive species ($Ar^+$, $C_2H_2^+$) as well as those for $C_4H_2^+$ and $C_4H_4^+$, are shown in figure 2, for constant bias (450 V) and pressure (12 mTorr). All species show a similar energy distribution with a dominant single peak at around 84 V. Secondary peaks are detectable at lower energies of 50 eV – 60 eV, with 1 – 2 orders of magnitude lower count rates. The position of the main peak is indicative of the positive potential developed between plasma and the chamber wall and varies little with substrate bias. For example, the $Ar^+$ peak energy increased by ~15% for a bias reduction of 350 V. These IEDs therefore are not truly representative of the ion arrival energies at the substrate.

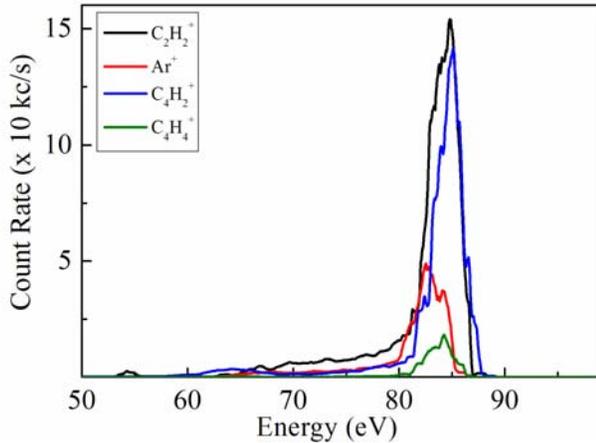

Figure 2 : Ion Energy Distributions of four dominant species measured at the wall in the CCP: 12 mTorr $C_2H_2$:Ar (flow ratio 2:1), $V_{dc}$ = 450 V at 13.56 MHz.

The ICP configuration allows the spectrometer head to be incorporated into the substrate electrode and driven at the same bias voltage and frequency. For a fixed ICP input power of 200 W at 13.56 MHz, with constant bias frequency (8.311 MHz) and pressure (3.3 mTorr), the dominant positive ions, $Ar^+$, $ArH^+$ $C_2H_2^+$ and $C_4H_2^+$ account for ~91.5% of ion flux across the energy range 0 – 200 eV. Figure 3 shows the IEDs of each of these species for a range of negative DC bias voltages up to 100 V. Count rates have been normalised to the value for $Ar^+$ at that bias and bias voltage sets are shifted vertically for clarity. Also shown are IEDs for a number of other species ($C_2H^+$, $CH_3^+$, $C_2H_4^+$) with normalised count rates scaled by a factor of five. For each species a bimodal energy peak structure is observed, with a mean energy that increases with bias and an energy width ($\Delta E$) that decreases with increasing mass. For each bimodal distribution, the ratio (I) of high energy to low energy peak heights is < 1 except for the lowest bias (-7 V). The $Ar^+$ IED also contained an additional broad peak at low energy (5 – 25 eV), the exact position of which depends on bias voltage.

In figure 4, IEDs for the same species are again plotted for varying bias, in this case for a high bias frequency (27 MHz) and similar characteristics are observed but with a significant reduction in $\Delta E$. At higher pressure (25 mTorr), the bimodal peaked structure is preserved, figure 5. However the dominant ion is now $ArH^+$, mainly at the expense of the $Ar^+$. The effect of changing the argon-acetylene flow ratio on the IEDs of the $Ar^+$ and $C_2H_2^+$ ion species is shown in figure 6. The data is obtained at low bias voltage and at a constant bias frequency (8.311 MHz) and pressure (3.3 mTorr). For both ion species, increasing the Ar:$C_2H_2$ flow ratio results in an increase in $\Delta E$, mainly due to a decrease in the position of the low energy peak. The peak height ratio, I, also decreases such that at 100% Ar, the IED consists mainly of a single low energy peak.

IEDs have also been obtained at pressures (20 mTorr and 120 mTorr) for zero and high bias values and are shown in figure 7. The figure shows only the $Ar^+$ ion but similar IEDs are obtained for hydrocarbon ions with greatly reduced flux.

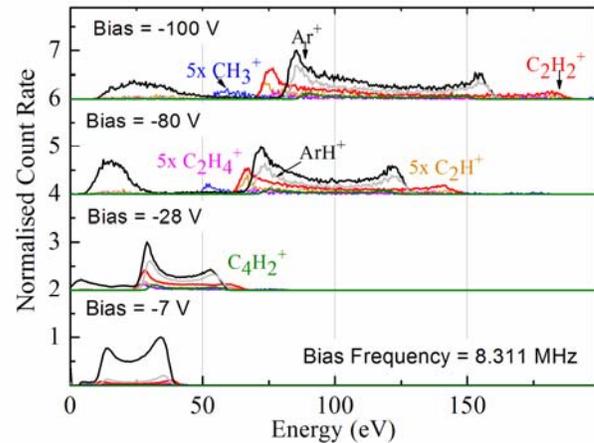

Figure 3 : Ion Energy Distributions of seven dominant species at the biased electrode of the ICP: 3.3 mTorr $C_2H_2$:Ar (flow ratio 2:1) $f_{bias}$ = 8.311 MHz and $V_{bias}$ varied. For clarity some ion fluxes are multiplied by 5 and shifted vertically.

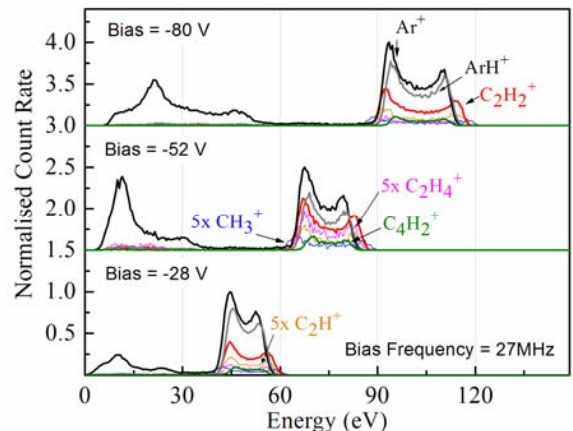

Figure 4 : Ion Energy Distributions of seven dominant species at the biased electrode of the ICP: 3.3 mTorr $C_2H_2$:Ar (flow ratio 2:1) $f_{bias}$ = 27 MHz and $V_{bias}$ varied. For clarity some ion fluxes are multiplied by 5 and shifted vertically.

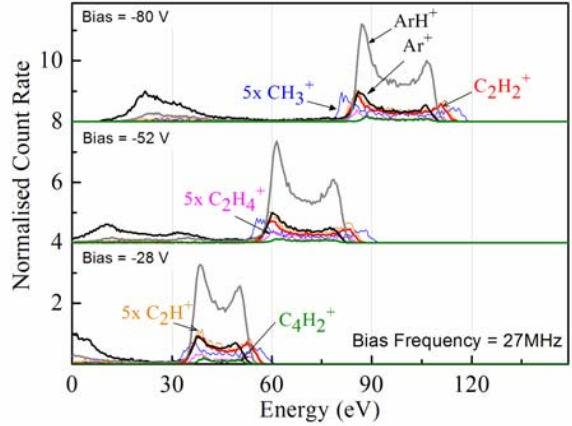

Figure 5 : Ion Energy Distributions of seven dominant species at the biased electrode of the ICP: 25 mTorr $C_2H_2$:Ar (flow ratio 2:1) $f_{bias}$ = 27 MHz and $V_{bias}$ varied. For clarity some ion fluxes are multiplied by 5 and shifted vertically.

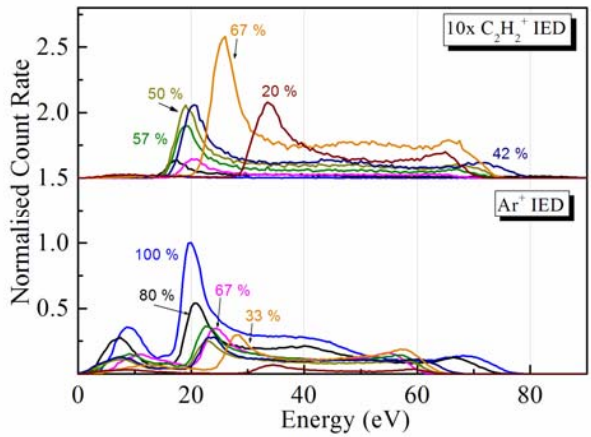

Figure 6 : Ion Energy Distributions of $Ar^+$ and $C_2H_2^+$ for varying argon fractions (marked on figure) measured at the biased electrode of the ICP: 3.3 mTorr $C_2H_2$:Ar (flow ratio 2:1) $f_{bias}$ = 8.311 MHz and $V_{bias}$ = 28 V. $C_2H_2^+$ IED vertically shifted and scaled 10 times for clarity.

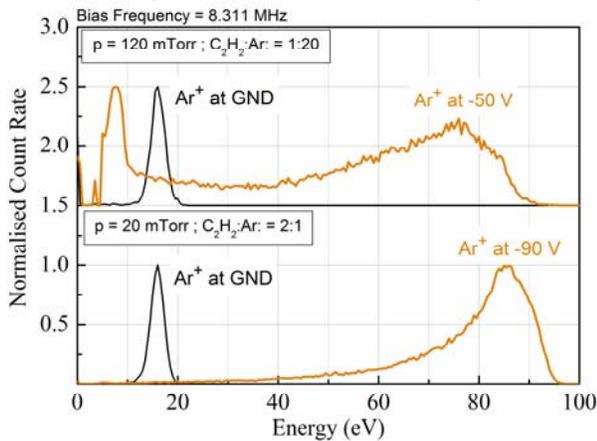

Figure 7 : Ion Energy Distributions of $Ar^+$ ion measured at the biased electrode of the ICP for two pressures and $V_{bias}$ varied at $f_{bias}$ = 8.311 MHz with different flow ratios of $C_2H_2$:Ar. For clarity IEDs for 120 mTorr are shifted vertically upward.

### 3.2 Film Deposition

Mass and ion energy spectrometry has been carried out in our film deposition chambers and in this section we present results on hydrogenated amorphous carbon (a-C:H) film characteristics deposited under similar process conditions. PECVD of a-C:H films is most commonly carried out in standard parallel plate CCP systems and hence we simply report relevant results carried out in the same chamber as the spectrometry, from our previous publications [5-9]. It should be noted however that the chamber required modification to accommodate the spectrometer head, resulting in an increase in the cathode – anode separation from 6 cm (deposition) to 16 cm. A number of films deposited under the latter configuration underwent rudimentary analysis and were found to be similar in quality to those obtained with the normal configuration. ICP deposition of a-C:H has received only limited attention to date and since in this configuration substrate bias variation is possible independent of plasma generation, we concentrate on this aspect of film deposition. Since the spectrometer head is incorporated within the substrate electrode, species measurement and deposition conditions are the same. However simultaneous measurement and deposition was not undertaken in order to control film thickness and minimise carbon loading of the internal spectrometer electrodes. Thus the orifice was closed during deposition. Determination of a-C:H film structure ($sp^3/sp^2$ bonding and hydrogen content) and properties involves a number of detailed high resolution materials analysis techniques. For this work however we are interested in relative changes in structure and hence we use features of the measured Raman spectra as indicators. We also made hardness measurements using nanoindentation at an approximately constant film thickness in order to obtain more direct indication of relative film quality under different conditions. More detailed materials analysis and film optimisation will be the subject of a future publication. Finally, we compare these results for a-C:H with our previously published work on hydrogen-free amorphous carbon films deposited via filtered cathodic vacuum arc and unbalanced magnetron sputtering techniques.

For a deposition time of 60 seconds, figure 8 shows the resultant film thickness variation with substrate bias for constant bias frequency (8.311 MHz), pressure (3.3 mTorr), gas flow ratio ($C_2H_2$:Ar = 2:1) and ICP power (200 W). The thickness is seen to drop significantly for bias voltages over 40 V.

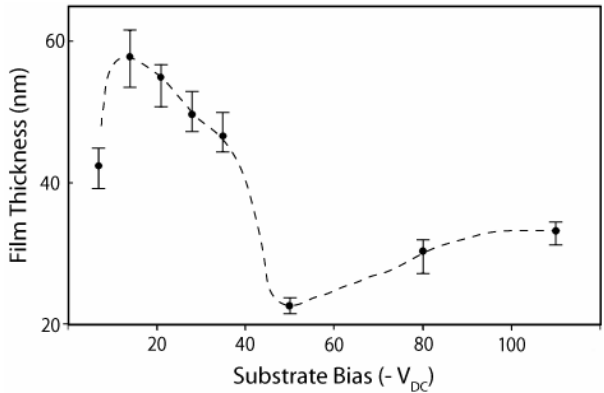

Figure 8 : DLC film thickness vs substrate bias in the ICP: 3.3 mTorr $C_2H_2$:Ar = 2:1, $f_{bias}$ = 8.311 MHz.

In figure 9, the Raman spectra for each film shows the characteristic peak at ~ 1550 cm$^{-1}$ (G-peak). Double Gaussian deconvolution of the spectra enables detection of a second peak at approximately 1380 cm$^{-1}$ (D-peak) from which the ratio of peak heights $I_D/I_G$ can be determined. This ratio is a useful indicator of the sp$^2$/sp$^3$ carbon-carbon bonding ratio provided the hydrogen content is relatively low. Ratios of 0.4 and lower are indicative of hard diamond-like films. Figure 10 shows the $I_D/I_G$ ratio to be ~ 0.4 for all films except for the highest bias. For soft films with a large hydrogen content, however, a low $I_D/I_G$ ratio can also be observed. Hydrogen content is difficult to estimate but it can give rise to significant luminescence under laser excitation and results in a noticeable slope in the Raman spectra [10]. Figure 11 shows the baseline slope of the spectra of figure 9, calculated in the regions from 500 cm$^{-1}$–1000 cm$^{-1}$ and 1700 cm$^{-1}$–2200 cm$^{-1}$. At low biases the slope is high but falls to almost zero for biases above ~ 40 V. A high hydrogen content at low bias is consistent with the greater growth rate observed under these conditions as the film density can be expected to be lower.

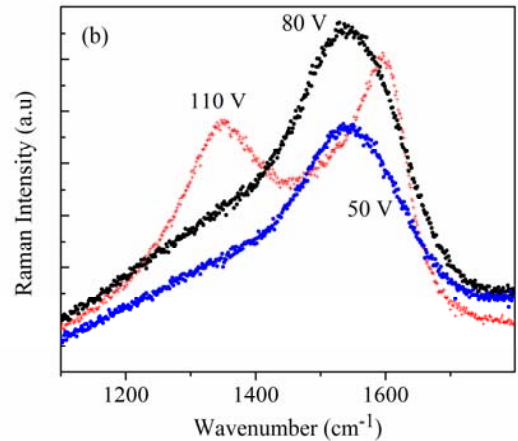

Figure 9 : Raman Spectra obtained for samples at each bias deposited in the ICP: 3.3 mTorr, $C_2H_2$:Ar = 2:1, $f_{bias}$ = 8.311 MHz. (a) $V_{bias}$ (7 V to 35 V) (b) $V_{bias}$ (50 V to 100 V).

Film hardness was obtained from nanoindentation measurements and is also shown in figure 10. Here we observe a clear peak in hardness at a bias of 80 V. It should be noted that for very thin films (~ 30 – 60 nm), the extracted hardness value depends on a number of factors such as penetration depth, tip blunting and the influence of the underlying substrate. For these measurements we used a fixed nanoindentation protocol and a constant penetration depth throughout in order to ensure that relative hardness values are reliable. More detailed analysis is currently in progress.

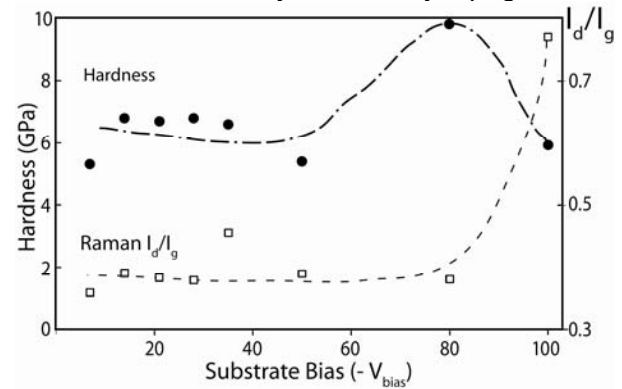

Figure 10 : Hardness and $I_d/I_g$ ratio vs substrate bias for the DLC samples deposited in the ICP: 3.3 mTorr $C_2H_2$:Ar = 2:1, $f_{bias}$ = 8.311 MHz.

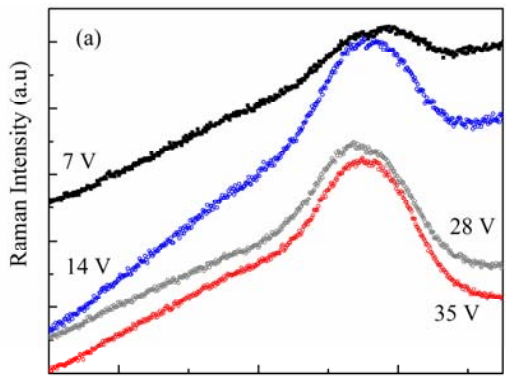

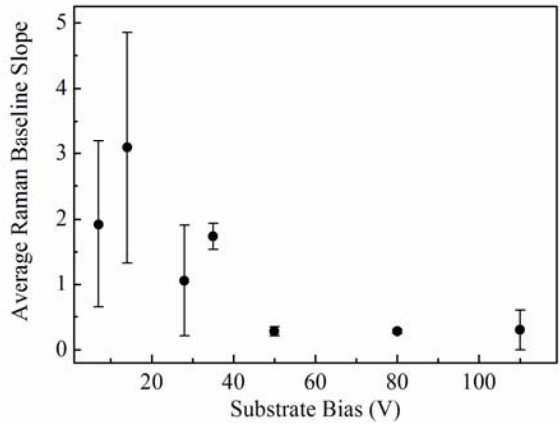

Figure 11 : Slope of Raman spectra extracted using graph depicted in figure 9.

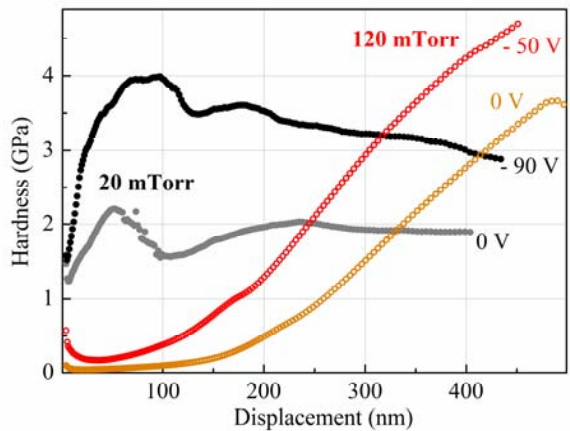

Figure 12 : Hardness vs displacement for the DLC films in the ICP: 20 mTorr $C_2H_2$:Ar = 2:1, $V_{bias}$ varied $f_{bias}$ = 8.311 MHz and 120 mTorr $C_2H_2$:Ar = 1:20, $V_{bias}$ varied $f_{bias}$ = 8.311 MHz.

Hardness measurements were also obtained from films deposited at pressures of 20 mTorr and 120 mTorr for both zero and high bias values of 50 V and 90 V respectively, figure 12, and it can be seen that the influence of the bias is still evident. The actual hardness is sensitive to tip displacement into the film and overall, as expected, these films are softer than those obtained at 3.3 mTorr, with a much higher growth rate and lower density. Nevertheless these films are still much harder than thin polymer-like films.

Finally, the properties of the ICP deposited films are compared with our previous results for CCP deposited a-C:H [5-9] and with hydrogen-free films t-aC [10, 11] and bias sputtered films [12, 13], figure 13. $I_D/I_G$ ratios are not directly comparable across hydrogenated and non-hydrogenated films and therefore the ICP ratios are equated to $sp^3$ fraction using conversion factors determined previously from detailed analysis of CCP films. The influence of substrate bias on the $sp^3$-fraction shows a similar trend for ICP a-C:H and non-hydrogenated samples, with a reduction beyond 50 – 100 V, whereas for CCP samples, the maximum $sp^3$ fraction is not obtained below biases of ~200 V, whereupon it remains constant up to ~500 V. Note that the hydrogen contribution to $sp^3$ content has not been excluded in the case of ICP samples and hence the carbon $sp^3$ fraction is likely to be overstated, particularly at lower biases where the films are softer.

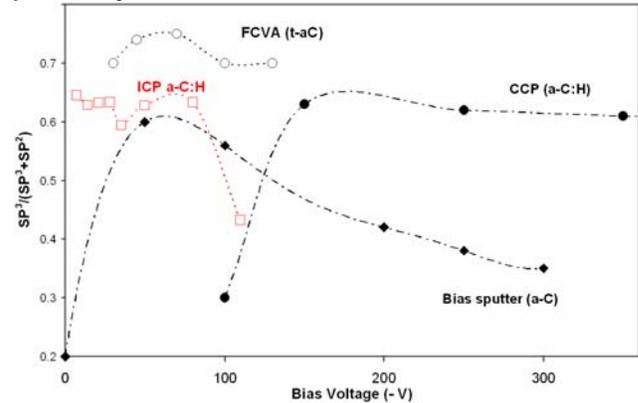

Figure 13 : Comparison of estimated $sp^3$ bonding ratio in ICP-deposited DLC (a-C:H) samples with films previously deposited via (i) CCP (a-c:H), in refs [5-9], and (ii) non-hydrogenated DLC deposited via Filtered Cathodic Vacuum Arc (FCVA), in refs [10-11], and via biased unbalanced magnetron sputtering, in refs [12-13].

## 4. Discussion
### 4.1 Ion Energy Distribution

We have measured IEDs for both CCP and ICP configurations. In the former the distribution, as measured on the grounded chamber wall, is dominated by a single peak for each of the four main species, $C_2H_2^+$, $C_4H_2^+$, $Ar^+$, $C_4H_4^+$, centred around 84 eV for a substrate bias of 450 V. The dominant ion is $C_2H_2^+$, most likely formed by direct ionisation of the parent molecule (11.4 eV) rather than charge exchange with $Ar^+$ ($10^{-16}$ $m^3$ $s^{-1}$) [14]. The $C_4H_2^+$ ion displays a similar level of flux to $C_2H_2^+$. In ICP generated IEDs we observe, at low pressure and mid-frequency (8.311 MHz), a different ordering of ion species with $Ar^+$ being dominant, followed by $ArH^+$ then $C_2H_2^+$. The flux of radical ions $C_2H^+$ and $CH_3^+$ have greater prominence compared to the CCP case, although still 1 – 2 orders of magnitude lower than $Ar^+$. The flux of $C_4H_2^+$ is now almost negligible.

In an associated paper [4], we observed that one of the dominant processes in $C_2H_2$:Ar plasmas involves hydrogen abstraction from the parent $C_2H_2$ molecule, via electron dissociative bombardment. This results in a pressure drop and a reduction in the set $C_2H_2$:Ar flow ratio. In general, an ICP is known to have electron densities typically 1 – 2 orders of magnitude greater than CCPs for similar power inputs. Furthermore, electron energy distribution functions (EEDF) measured in the ICP indicate a greater proportion of electrons of intermediate energy (3eV- 18eV) [4, 15] compared to the bi-Maxwellian EEDFs expected in CCPs. Thus, in a pure argon plasma the ICP configuration has the greater $Ar^+$ density, compared to CCP, for equivalent power density. In the $C_2H_2$:Ar ICP plasma conditions therefore favour electron dissociation over ionisation of acetylene to a much greater degree than for the CCP. Thus the lower observed $C_2H_2^+$ ion flux in the ICP is to be expected.

From a materials deposition perspective, the nature of the ion flux is important. Current models [1] assume a dominant carbon-carrying ion flux which contributes to growth via direct incorporation of carbon (as well as hydrogen) ions into the growing film. The energy of the carbon ions is thought to be important in determining the final bonding ($sp^3/sp^2$) and hydrogen content. For ICP deposited films, however, the flux of carbon-carrying ions is extremely small and the energetics of growth is therefore dominated by inert argon (or $ArH^+$) bombardment.

Growth of hydrocarbon films is a complex function of both synergistic and competing reactions. In its simplest form, this growth involves (i) direct incorporation of radical dissociation products of the precursor ($C_2H_2$), which depends on their sticking coefficient, (ii) direct incorporation of secondary reaction products of relatively high sticking coefficient, particularly the higher hydrocarbons, (iii) indirect incorporation of primary dissociation or secondary reaction products via surface dangling bond activation involving ion-bombardment and/or atomic hydrogen reactions, (iv) direct incorporation of energetic carbonaceous ions and (v) reduction of film hydrogen content via ion bombardment or film etching by atomic hydrogen. Investigation of the dominance of the different reaction channels is hampered by the limited data on plasma species, flux and energies involved. A number of dedicated experiments, for example, involving pure radical beams [2], pulsed plasma [16] and remote ECR plasma sources [1] have been undertaken to elucidate the various reaction channels. These have concentrated mainly on the radical-based channels while ion-bombardment reactions, especially at energies normally found in plasma deposition, have received much less attention. In this work, we observe a distinct hardness-bias relationship for the ICP films where $Ar^+/C_2H_2^+ >> 1$ while for CCP films, where $Ar^+/C_2H_2^+ << 1$, the bias voltages are much greater (> 150 V) and the bias dependence is minimal above a threshold. However, such a straightforward comparison does not account for the different distribution (IED) of ion energies at the substrate. In terms of IED shape, we observe a distinct difference between the single-peaked structures obtained from the CCP compared to the bimodal peaked ICP distributions. In the latter case, a third low energy $Ar^+$ peak was often observed, possibly due to collisions, while the exact features of the bimodal structure depended on bias and frequency. Since the CCP measurement is made at the wall, however, it is not clear whether the single peak is truly representative of the IED at the substrate.

The shape of an IED, from simple analytical models, is expected to exhibit bimodal peaks with the width of the distribution varying as a function of ion mass, rf voltage and frequency. At sufficiently high frequency, the ratio of ion transit time to rf period is long and the ion sees effectively a DC potential, leading to a single-peak IED. $Ar^+$ IED models by Economou [17] show a double peak at 13.56 MHz, while both models and measurements by Sobolewski [18] show a single peak distribution only at higher frequencies (30 MHz) for $CF_4$ derived species in an ICP discharge. The latter was measured at the grounded chamber wall and the observed IED was related to that at the driven electrode through bias-induced variation in the plasma potential. While these models and results relate particularly to high density ICP systems, other reports for CCP operation also show bimodal distributions [19, 20] for IEDs measured at the driven substrate. However, IED measurements by Tatsuta [21] under rf bias conditions, showed, for both $Ar^+$ and $CH_4^+$ CCP plasmas, a dominant single peak situated at energies $\geq V_{bias}$ and a smaller "collisional" peak at lower energies. For similar measurements at the ground electrode, this single peak was less pronounced, while the "collisional" peak had broadened. For a 13.56 MHz $Ar/CH_4$ plasma, Jie et al [15] observed single peak IED characteristics as did Jacob [1]. These were obtained at a grounded electrode remote from an ICP or ECR source, respectively.

An IED observed at the grounded chamber wall is determined by the bias-induced variation in plasma potential and will be related to the true IED at the biased electrode through the ratio of electrode sheath impedance to wall sheath impedance ($Z_e/Z_w$). When the capacitive component of impedance dominates (as is the case in a CCP), then $Z_w << Z_e$, due to the much larger wall area, and the wall sheath voltage is a small fraction of the rf bias voltage. This factor may explain our observed peak energies of ~ 80 V which are much less than the DC bias (450 V). Also, measurements of the RF component of the plasma potential in a similar CCP system [22] show magnitudes around 10 V, similar to the FWHM of the IED peaks. It is reasonable to assume therefore that the true substrate IED in the CCP case is not represented by the IED measured at the wall. Instead, using published data and models [19, 23], we would expect a pure bimodal peaked distribution centred at $V_{bias}$ and with an energy spread of $kV_{bias}$, where k ~ 0.1 – 0.3. However this applies only for a truly collisionless plasma (p < 1 mTorr) and for pressures around 7 mTorr we could expect a flat energy distribution from zero to $V_{bias}$ with numerous superimposed peaks, due to sheath collisions [23]. As the bias is reduced, the background distribution becomes more asymmetric, with the lower energies dominant. A similar trend is described by Manenschijn [19]. These peaks have been attributed to charge exchange collisions while elastic ion-neutral scattering results in a broad, smooth background distribution.

The shape of the measured ICP IEDs, within the bimodal region (i.e. excluding lower energy collisional peaks), can be characterised by a number of related parameters, namely centre energy ($E_{mid}$), mean energy ($E_{mean}$), energy width (FWHM) ($\Delta E$), high and low energy peak positions ($E_H$, $E_L$) and amplitude ratio (I) for ion counts at $E_H/E_L$. From a film growth perspective, we ideally require a small value of $\Delta E$ and a high value of I, tuneable over the energy range, as this allows a definitive exploration of ion energy effect on growth models. $\Delta E$ is determined by the rf bias amplitude at low frequency but at very high frequencies ions traverse the sheath over multiple RF periods and 'see' only an average sheath potential. A number of analytical expressions have been derived [18] which relate $\Delta E$ against ion mass and frequency, namely:

$$\frac{\Delta E}{V_{pp}} = \frac{2e}{\pi}\left(\frac{\tau_{ion}}{T}\right)^{-1} \qquad (1)$$

and

$$\frac{\Delta E}{V_{pp}} = e\left(1 + \left(\frac{2.25\tau_{ion}}{T}\right)^2\right)^{-1} \quad (2)$$

where, $V_{pp}$ is the peak-to-peak rf bias voltage and $\tau_{ion}/T$ is the ratio of the ion transit time to the rf period, given by

$$\frac{\tau_{ion}}{T} = 2\left(\frac{V_{pp}}{2em^*}\right)^{\frac{1}{4}}\left(\frac{m_{ion}\varepsilon_0}{J_+}\right)^{\frac{1}{2}} f \quad (3)$$

where $m^*$ is the weighted average ion mass, $J_+$ the total ion flux, $m_{ion}$ the ion mass and f the frequency. The second model (B), equation (2), accounts for the low frequency situation where ions can follow the slowly changing voltage and $\Delta E/V_{pp} = 1$.

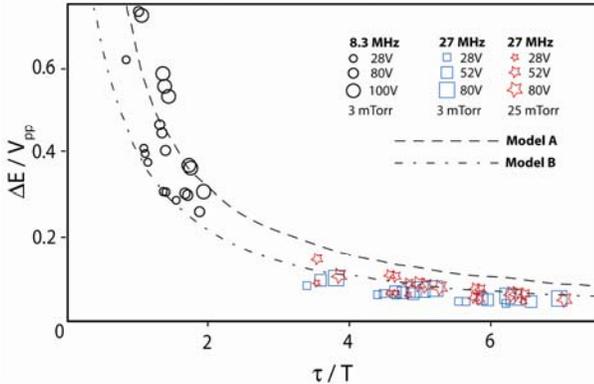

Figure 14 : Measured (symbols) and model [18] (dotted curves) values of $\Delta E/V_{pp}$, the normalized separation between the two peaks in the ion energy distributions, as a function of the ratio of ion transit time $\tau$ to rf time period T at an ICP input power of 200 W at pressures of 3.3 mTorr and 25 mTorr for varying $V_{bias}$ and $f_{bias}$.

In figure 14, we plot measured $\Delta E/V_{pp}$ against $\tau_{ion}/T$ for 8.311 MHz and 27 MHz for 3.3 mTorr and 25 mTorr, and compare with model A, equation (1) and model B, equation (2). We used an estimate of the total flux from flux probe and Langmuir probe measurements and $V_{pp}$ was obtained from voltage measurement near the electrode, adjusted to account for transmission line effects, to which was added the value of plasma potential (~20V). Since the waveform contains significant harmonic content, the electrical distance between the measure point and the electrode needs to be considered as a transmission line, in order to obtain the true voltage. We have derived an equivalent electrical model for this system and estimate that the actual electrode voltage can be up to 30% higher. At high frequencies, $\Delta E/V_{pp}$ is very small (~ 0.1) and follows closely model B, whereas at 8.311 MHz, the energy width is much greater and the data lies closer to model A, which does not account for the low frequency situation. However, errors in determining sheath potential difference must be considered here as the measured bias waveforms sometimes show a high harmonic content. Also the plasma impedance and the rf component of the plasma potential also need to be considered and these can vary considerably with plasma conditions [24].

The ICP bimodal IED structures are well developed, even at relatively high pressures of 25 mTorr, where in the CCP case collisional effects are reported to be dominant. The likely reason for this is the much smaller sheath width in an ICP. In our case, average sheath widths are estimated, optically, to be around 0.5 mm compared to reported CCP values of 5 mm – 10 mm. Nevertheless some broad low energy peaks are observed at low and high frequency for $Ar^+$ species and it is reasonable to attribute these to ion-neutral collisions within the sheath. For ions travelling in their parent gas, the charge exchange cross-section is normally greater than that for elastic collisions [19], hence the presence of $Ar^+$ collision peaks and the absence of $C_xH_y^+$ peaks is not unexpected. At higher pressure, 25 mTorr, we observe a reduction in the height of the $Ar^+$ collisional peaks. However the dominant ion is now $ArH^+$ and charge exchange between it and the Ar gas may therefore be less probable. In summary therefore we expect CCP IEDs to be relatively broad and bias independent while ICP energy distributions depend on the structure of the bimodal peaks which can be tailored to an extent by plasma input parameters, namely; bias and frequency.

One, rarely considered, aspect of the bimodal structure is the ratio of the high and low peak heights (I) which impact on the average ion energy. Highly asymmetric distributions can be considered equivalent to monoenergetic beams, ideal for film growth modelling. For all conditions, except for the lowest bias (7 V) at 8.311 MHz, the ratio I is less than 1. At 27 MHz, it is approximately constant (0.5 – 0.75) across bias and pressure while for 8.311 MHz it tends to increase with bias, figure 15.

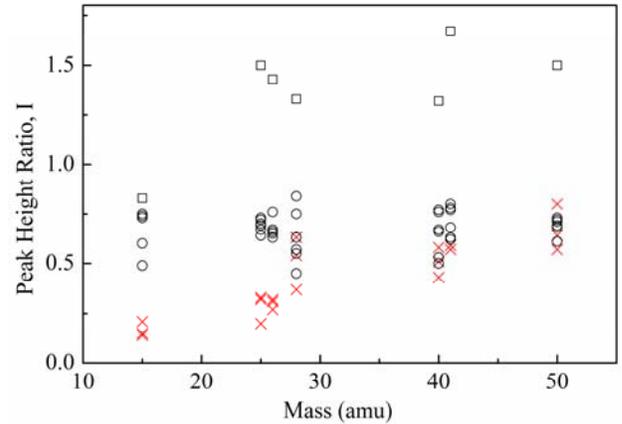

Figure 15: IED peak height ratio vs mass for the seven dominant species. Squares: $V_{bias}$ = 7 V $f_{bias}$ = 8.311 MHz. Crosses $V_{bias}$ > 28 V $f_{bias}$ = 8.311 MHz. Circles: $V_{bias}$ > 28 V $f_{bias}$ = 27 MHz.

To date, little attention has been paid to this characteristic in the literature. We have constructed a simple numerical ion track model to investigate the factors influencing the ratio I using a RF sheath model [25] based on a single-species Child-Langmuir sheath under rf excitation. The model inputs are the measured electron temperature ($T_e$) and density ($n_e$) from [4] and $V_{pe}$, the time-varying potential across the sheath determined from rf voltage measurements close to the electrode and the plasma potential $V_{pl}$. The rf period is divided into 1200 time steps and single ions are injected into the sheath at each time step, with a velocity derived from the Bohm criterion [25]. The sheath potential variation is divided into 500 equal distance steps and an estimate of average sheath width is obtained from optical measurements and then adjusted to obtain the best fit for the low energy cut-off point in the bimodal part of the IED. The applied voltage is assumed sinusoidal and the small rf ripple on $V_{pl}$ is ignored.

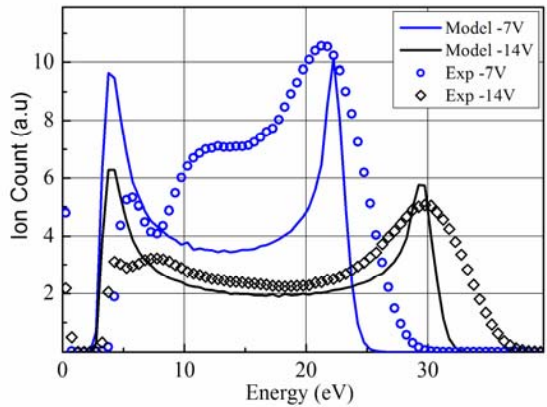

Figure 16 : One dimensional Ion track Model IED (line) obtained for two $V_{bias}$ at $f_{bias}$ = 8.311 MHz using pure sinusoidal bias waveform compared with the measured IED (marker): 3.3 mTorr $C_2H_2$:Ar (flow ratio 2:1) and same substrate bias condition.

Figures 16 show a good fit between the model and measured data at low biases of 7 V and 14 V, although the measurement exhibits a double peak in the low energy region. At higher biases the model fit is much less reliable. This is partly due to discretisation errors, given the model simplicity. However much more critical is the shape of the input bias waveform. In figure 17, the model output is given for a pure sinusoidal input at 85 V bias (8.311 MHz) and the peak ratio is ~1 as observed with the measured data. In figures 18, two different harmonically rich waveforms are input and the resulting IEDs peak ratios show extreme values of I >> 1 or I ~ 0.

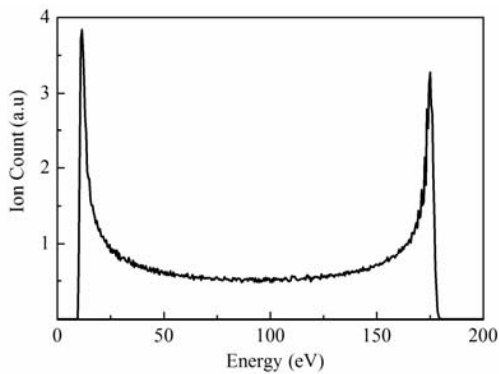

Figure 17: Modelled IED, $V_{bias}$ = 85 V pure sinusoid. 1200 ions, Gaussian velocity distribution assumed. Peak ratio ~ 1 (almost symmetric).

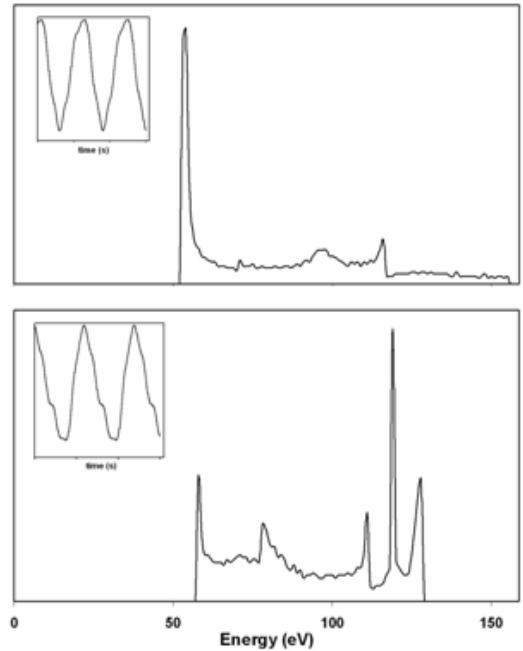

Figure 18 : One dimensional Ion track modelled IEDs for two different harmonically rich waveforms.

While ICP processing offers the opportunity for greater control of ion energy range compared to CCP, realising a highly monoenergetic ion beam would be valuable for control of film growth. One approach, suggested by these results, is to use high frequency biasing. Alternatively, the observed sensitivity to waveform harmonic content offers the possibility of IED tailoring using multi-frequency biasing.

**4.2 Film deposition**

A small number of films were deposited at a single bias frequency (8.311 MHz) in order to confirm that the conditions for plasma species measurements were suitable for film growth. No attempt was made at this stage to optimise growth conditions for adhesion, thickness, hardness etc. The major difference between films deposited by ICP and CCP is the peaked response in the hardness – bias curve for the ICP case. To account for the spread in ion energy, the hardness data from figure 10 is replotted, in figure 19, against mean ion energy ($\bar{E}$) and energy spread ($\Delta E$). At low biases the softer films are consistent with a relatively high hydrogen content, as evidenced by the Raman baseline slope, figure 11, and the higher growth rate, figure 8, which implies a reduced density since the arriving species flux varies little across the bias range. Similar quality films are often obtained at the grounded electrode of a CCP. On the high bias side, the soft films exhibit low hydrogen content and are more graphite-like, as found with CCP films deposited under very high bias. It is expected that high energy ion bombardment will cause hydrogen depletion and allow the remaining C atoms to re-arrange via $sp^2$-bonding. The striking feature here, however, is the relatively low value of energy required and the appearance of a very sharp threshold between 90 – 100 eV.

The ICP hardness characteristic is similar to that obtained for our hydrogen-free tetragonal amorphous carbon (t-aC), figure 13. However t-aC is formed by (mainly) $C^+$ ion bombardment and the high $sp^3$-fraction and hardness values arise from sub-surface carbon implantation in a restricted volume. Surface penetration and atom displacement energy thresholds must be overcome to achieve $sp^3$-bonding otherwise low energy ions are simply adsorbed on the surface in $sp^2$-configuration. However, excess ion energy causes thermal relaxation of the $sp^3$-bonds to the natural $sp^2$-state. In the ICP case, however, we can consider the ion flux to consist of only $Ar^+$ ions and hence the t-aC model is only appropriate for implantation via $Ar^+$ induced knock-on of surface adsorbed carbon species. The ICP hardness characteristic also shares some similarity with that of hydrogen-free sputtered films [12, 13], figure 13. Normally, sputtered films are relatively soft and graphite-like ($sp^2 > 80\%$) since the carbon flux has low energy. With an unbalanced magnetron configuration and added substrate bias, the substrate ion flux and energy are enhanced such that the adsorbed carbon atoms are bombarded by $Ar^+$ ions, allowing knock-on implantation to increase the $sp^3$-bonding. The resultant hardness maximum occurs around 50 V (negative) bias and falls off slowly thereafter. Calculations for t-aC [26] show a maximum implantation effect at 90 eV per C atom and an effective penetration threshold of ~ 44 eV. However, the thermal relaxation process occurs over a > 50 eV "range" above the peak energy, much larger than that observed here (< 10 eV). A similar characteristic was obtained from an implantation model for sputtered films [27] with a peak at energies around 80 eV. However the model used arbitrary fitting parameters to match the width of the characteristic to experimental sputtered values. Reconciliation of this relaxation factor with the ICP results may be possible by considering additional effects due to higher energy ions (> 130 eV) present under the 100 V bias conditions, given the greater $\Delta E$. However, their flux is low and such a consideration would be more relevant to the CCP films and hence can be discounted. In fact the wide bias window (150 V – 500 V) for high $sp^3$-fraction CCP films is consistent with the assumption of a collisional IED with a broad energy range up to the bias value. The dominance of the $C_2H_2^+$ ion in this case should also lead to direct carbon implantation. Other ion bombardment effects, i.e. surface hydrogen removal, creation of surface dangling bonds, sub-surface reorganisation and H removal must also be involved in film formation although their significance in comparison with sub-surface implantation cannot be evaluated. It should be noted that high hardness amorphous carbon films have been obtained [28] without significant ion flux or energy using an expanding thermal plasma to dissociate acetylene and create a high flux of $C_2H$ radicals. This reaction channel may be of some importance in ICP film formation due to the almost total $C_2H_2$ dissociation and high $C_2H$ radical production rate [4].

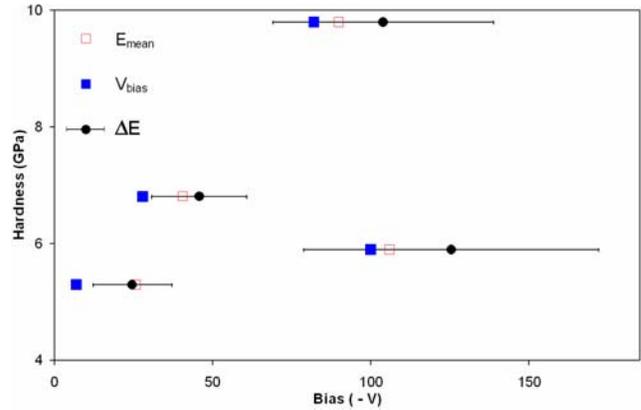

Figure 19: Hardness v bias curve from figure 10 replotted to also show dependence on average energy $E_{mean}$ (□) and $\Delta E$ of the ion flux. Data was calculated from the bimodal portions of the respective IEDs, ignoring collisional peaks.

Finally, the effect of pressure on film deposition was briefly investigated. For a set pressure of 60 mTorr, the working plasma pressure was 20 mTorr and the IED exhibited a single peak for $Ar^+$ at the bias voltage (or plasma potential for the unbiased case), figure 7. No lower energy collisional peaks were observed. The film hardness was found to be reduced by more than a factor of 2 compared to the low pressure case although the ion energy was at the optimum value ~ 90 eV. For higher pressures (120 mTorr working pressure) a low energy collisional peak is dominant and the hardness is further reduced. At these high pressures, the ion fluxes are significantly less than at low pressure. Nevertheless the effect of ion bombardment is still observable in the difference in hardness for biased and unbiased films, figure 12.

## 5. Conclusion

Ion energy distributions have been determined at the rf-driven electrode in an inductively-coupled acetylene-argon plasma for various substrate bias voltages and frequencies under conditions suitable for film deposition. The plasma ion content is predominantly $Ar^+$ or $ArH^+$ despite the set flow ratio of $C_2H_2$:Ar of 2:1 due to the high dissociation of the parent molecule [4]. For a range of pressures, up to 25 mTorr, the IEDs exhibit well developed bimodal structures with peak separation values that follow the expected dependence on voltage and frequency. Additional lower energy collisional peaks are also observed. At higher pressures, up to 120 mTorr, the bimodal structure is replaced by a single peak and low energy collisional peaks become dominant. Low pressure deposited diamond-like carbon films indicate a peak hardness at an ion energy of around 90 eV and a very sharp fall in hardness beyond this. This is similar to models of $sp^3$ bond formation in hydrogen-free tetragonal amorphous carbon or bias-sputtered films. However, due to the lack of carbon-based ions, the mechanism may be attributed to argon knock-on implantation of surface adsorbed carbon species. Other, complementary ion- and radical-based growth mechanisms are also likely to be involved. While the direct comparison of ion energy distributions and film characteristics has provided useful insight into growth mechanisms, greater control of the IED width would be beneficial in providing almost monoenergetic ion beams under true plasma deposition conditions. Results from this work have shown that use of high frequency bias or bias harmonics may lead to much narrower distributions and film deposition under these conditions is currently under investigation.